# Interference Channels with Common Information

## Jinhua Jiang, Yan Xin, and Hari Krishna Garg

### Abstract

In this paper, we consider the discrete memoryless interference channel with common information, in which two senders need deliver not only private messages but also certain common messages to their *corresponding* receivers. We derive an achievable rate region for such a channel by exploiting a random coding strategy, namely *cascaded* superposition coding. We reveal that the derived achievable rate region generalizes some important existing results for the interference channels with or without common information. Furthermore, we specialize to a class of deterministic interference channels with common information, and show that the derived achievable rate region is indeed the capacity region for this class of channels.

## I. INTRODUCTION

The interference channel (IC) is one of the fundamental building blocks in communication networks, in which the transmissions between each sender and its corresponding receiver (each sender-receiver pair) take place simultaneously and interfere with each other. The information-theoretic study of such a channel was initiated by Shannon in [1], and has been continued by many others [2]–[16] and etc. So far, the capacity region of the general IC remains unknown except for some special cases, such as the IC with strong interference (SIC) [3], [6], [9], [10], [12], a class of discrete additive degraded ICs [8], and a class of deterministic ICs [11]. Alternatively, various achievable rate regions served as inner bounds of the capacity region have been derived for the general IC [5], [7], [9], [15]. Notably, Carleial [7] obtained an achievable rate region of the discrete memoryless IC by employing a limited form of the general superposition coding scheme [17], *successive* encoding and decoding. Subsequently, Han and Kobayashi [9] established the best achievable rate region known till date by applying the *simultaneous* superposition coding scheme consisting of simultaneous encoding and decoding. Indeed, the improvement of the Han-Kobayashi region [9] over the Carleial region [7] is primarily due to the use of the simultaneous decoding. This has been validated in [15], [16], in which Chong *et al.* obtained an achievable rate

The authors are with the Department of Electrical and Computer Engineering, National University of Singapore, 4 Engineering Drive 3, Singapore, 117576, (e-mails: Jinhua.Jiang@nus.edu.sg, elexy@nus.edu.sg, and eleghk@nus.edu.sg). The corresponding author is Dr. Yan Xin, tel. (+65) 6516 5513, fax (+65) 6779 1103.

region identical with the Han-Kobayashi region but with a much simplified description, by using a hybrid of the successive encoding (same as Carleial's) and simultaneous decoding. Moreover, Carleial [7] introduced the notion of the partial cross-observability of each sender's private information, which means that each receiver is able to decode part of the private information sent from its non-pairing sender. The derivation of the Han-Kobayashi region and Chong-Motani-Garg region followed this notion but Chong *et al.* made an important observation that the decoding errors of the crossly observed information can be excluded in computing the probability of error [15]. With an introduction of the partial cross-observability, the IC can be viewed as a compound channel consisting of two associated multiple access channels (MACs) (strictly speaking, MAC-like channels), and thus its achievable rate region can be approached by exploiting existing techniques used for MACs. However, the proof of the converse for either achievable rate region (Han-Kobayashi region or Chong-Motani-Garg region) is still not available.

Most of the prior work on the ICs assumes the statistical independence of the source messages to be transmitted by the senders [2]–[16]. However, this assumption becomes invalid in an IC where the senders need transmit not only the private information but also certain common information to their corresponding receivers. Such a scenario is generally modelled as the IC with common information (ICC). Maric *et al.* [18] derived the capacity region of a special case of the ICC, the strong interference channel with common information (SICC), and their result subsumes the capacity region of the SIC (without common information) [12] as a special case. Parallel to the case of the IC, the study of the ICC is closely related to the prior work on the MAC with common information (MACC) that has been thoroughly studied by Slepian and Wolf in [19] and Willems in [20]. As an example, the capacity region of the SICC shown in [18] can be interpreted as an intersection of the capacity regions of two underlying MACC-like channels. Moreover, our main results also develop upon interpreting an ICC as a composite channel of two MACC-like channels.

In this paper, we propose a generalized version of the successive superposition encoding, namely *cascaded* superposition encoding, which reduces to Carleial's successive encoding in the absence of common information. With this encoding scheme, the senders' common information is conveyed through the channel in a cooperative manner. Applying the proposed cascaded encoding scheme along with the simultaneous decoding scheme [9], [15], we derive an achievable rate region for the *general* two-user discrete memoryless ICC. The derived achievable rate region subsumes the Chong-Motani-Garg region for the general IC as well as the capacity region for the SICC as special cases. Moreover, we derive an achievable rate region for a particular class of ICCs where one of the two senders has no private information for its corresponding receiver. The depiction of the obtained achievable rate region appears very simple with only one auxiliary random variable involved. Proving the converse still appears to be a challenge, which we believe is as difficult as proving the converse for the Han-Kobayashi region or the Chong-Motani-Garg region.

Lastly, we investigate a class of deterministic interference channels with common information (DICCs), which generalizes the class of DICs (without common information) studied in [11]. Relying on the crucial assumptions we specified for this class of channels, we show that our achievable rate region meets the outer bound of the capacity region, and thus it is actually the capacity region of this class channels. This in a certain sense indicates the potential tightness of the region as an inner bound of the capacity region of the general discrete memoryless ICC.

The rest of the paper is organized as follows. In Section II, we first introduce our channel models, including the general ICC and a modified ICC. The modified ICC serves to reveal the information flow through the associated ICC, and facilitates the derivation of the achievable rate region for the associated ICC. In Section III, we present the achievable rate region for the general discrete memoryless ICC in both implicit and explicit forms. We also provide a detailed proof of the achievability of the rate region. In Section V, we apply the obtained achievable rate region to three special cases of the ICC including the SICC, the general IC (without common information), and a class of the ICCs where one of the two senders has no private information to transmit. For each case, our achievable rate region either includes the existing results as special cases or gives a new achievable rate region. In Section IV, we investigate the class of DICCs for which our achievable rate region is in fact the capacity region. The paper is concluded in Section VI.

*Notations:* Random variables and their realizations are denoted by upper case letters and lower case letters respectively, e.g., $X$ and $x$. Bold fonts are used to indicate vectors, e.g., $\mathbf{X}$ and $\mathbf{x}$.

## II. CHANNEL MODELS AND PRELIMINARIES

### A. Discrete Memoryless Interference Channel with Common Information

A discrete memoryless IC is usually defined by a quintuple $(\mathcal{X}_1, \mathcal{X}_2, \mathcal{P}, \mathcal{Y}_1, \mathcal{Y}_2)$, where $\mathcal{X}_t$ and $\mathcal{Y}_t$, $t = 1, 2$, denote the finite channel input and output alphabets respectively, and $\mathcal{P}$ denotes the collection of the conditional probabilities $p(y_1, y_2 | x_1, x_2)$ of the receivers obtaining $(y_1, y_2) \in \mathcal{Y}_1 \times \mathcal{Y}_2$ given that $(x_1, x_2) \in \mathcal{X}_1 \times \mathcal{X}_2$ are transmitted. The channel is memoryless in the sense that for $n$ channel uses, we have

$$p(\mathbf{y}_1, \mathbf{y}_2 | \mathbf{x}_1, \mathbf{x}_2) = \prod_{i=1}^{n} p(y_{1i}, y_{2i} | x_{1i}, x_{2i}),$$

where $\mathbf{x}_t = (x_{t1}, ..., x_{tn}) \in \mathcal{X}_t^n$ and $\mathbf{y}_t = (y_{t1}, ..., y_{tn}) \in \mathcal{Y}_t^n$ for $t = 1, 2$. The marginal distributions of $y_1$ and $y_2$ are given by

$$p_1(y_1 | x_1, x_2) = \sum_{y_2 \in \mathcal{Y}_2} p(y_1, y_2 | x_1, x_2),$$

$$p_2(y_2 | x_1, x_2) = \sum_{y_1 \in \mathcal{Y}_1} p(y_1, y_2 | x_1, x_2).$$

Building upon an IC, we depict an ICC in Fig. 1. Sender $t$, $t = 1, 2$, is to send a private message $w_t \in \mathcal{M}_t = \{1, ..., M_t\}$ together with a common message $w_0 \in \mathcal{M}_0 = \{1, ..., M_0\}$ to its pairing receiver. All the three messages are assumed to be independently and uniformly generated over their respective ranges.

Let $C$ denote the discrete memoryless ICC defined above. An $(M_0, M_1, M_2, n, P_e)$ code exists for the channel $C$, if and only if there exist two encoding functions

$$f_1 : \mathcal{M}_0 \times \mathcal{M}_1 \to \mathcal{X}_1^n, \qquad\qquad f_2 : \mathcal{M}_0 \times \mathcal{M}_2 \to \mathcal{X}_2^n,$$

and two decoding functions

$$g_1 : \mathcal{Y}_1^n \to \mathcal{M}_0 \times \mathcal{M}_1, \qquad\qquad g_2 : \mathcal{Y}_2^n \to \mathcal{M}_0 \times \mathcal{M}_2,$$

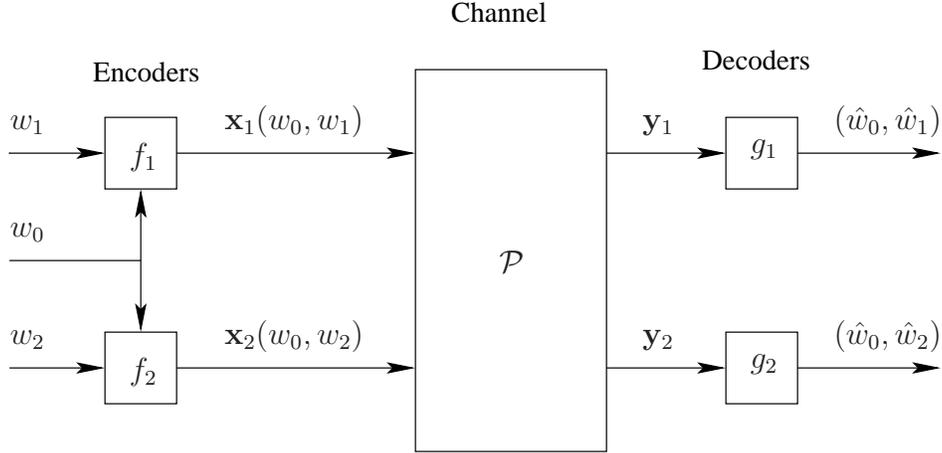

Fig. 1. The interference channel with common information.

such that $\max\{P_{e,1}^{(n)}, P_{e,2}^{(n)}\} \leq P_e$, where $P_{e,t}^{(n)}$, $t = 1, 2$, denotes the average decoding error probability of decoder $t$, and is computed by one of the following:

$$P_{e,1}^{(n)} = \frac{1}{M_0 M_1 M_2} \sum_{w_0 w_1 w_2} p((\hat{w}_0, \hat{w}_1) \neq (w_0, w_1)|(w_0, w_1, w_2)),$$

$$P_{e,2}^{(n)} = \frac{1}{M_0 M_1 M_2} \sum_{w_0 w_1 w_2} p((\hat{w}_0, \hat{w}_2) \neq (w_0, w_2)|(w_0, w_1, w_2)).$$

A non-negative rate triple $(R_0, R_1, R_2)$ is achievable for the channel $C$ if for any given $0 < P_e < 1$, and for any sufficiently large $n$, there exists a $(2^{nR_0}, 2^{nR_1}, 2^{nR_2}, n, P_e)$ code.

The capacity region for the channel $C$ is defined as the closure of the set of all the achievable rate triples, while an achievable rate region for the channel $C$ is a subset of the capacity region.

### B. Modified Interference Channel with Common Information

To derive an achievable rate region for the ICC, we first need to be clear about the structure of the information flow through it. However, this can not be viewed from the original ICC model clearly, and thus it is difficult for one to carry out the corresponding information-theoretic analysis. To avoid such difficulty, we introduce the modified ICC by following the same approach used in [9].

The modified ICC inherits the same channel characteristics from its associated ICC, but there are five streams of messages to be conveyed through the modified channel instead of three through the associated ICC. The five streams of messages $n_0$, $n_1$, $l_1$, $n_2$ and $l_1$ are assumed to be independently and uniformly generated over the finite sets $\mathcal{N}_0 = \{1, ..., N_0\}$, $\mathcal{N}_1 = \{1, ..., N_1\}$, $\mathcal{L}_1 = \{1, ..., L_1\}$, $\mathcal{N}_2 = \{1, ..., N_2\}$, and $\mathcal{L}_2 = \{1, ..., L_2\}$, respectively.

Denote the modified ICC shown in Fig. 2 by the channel $C_m$. An $(N_0, N_1, L_1, N_2, L_2, n, P_e)$ code exists for $C_m$ if and only if there exist two encoding functions

$$f_1 : \mathcal{N}_0 \times \mathcal{N}_1 \times \mathcal{L}_1 \to \mathcal{X}_1^n, \qquad f_2 : \mathcal{N}_0 \times \mathcal{N}_2 \times \mathcal{L}_2 \to \mathcal{X}_2^n,$$

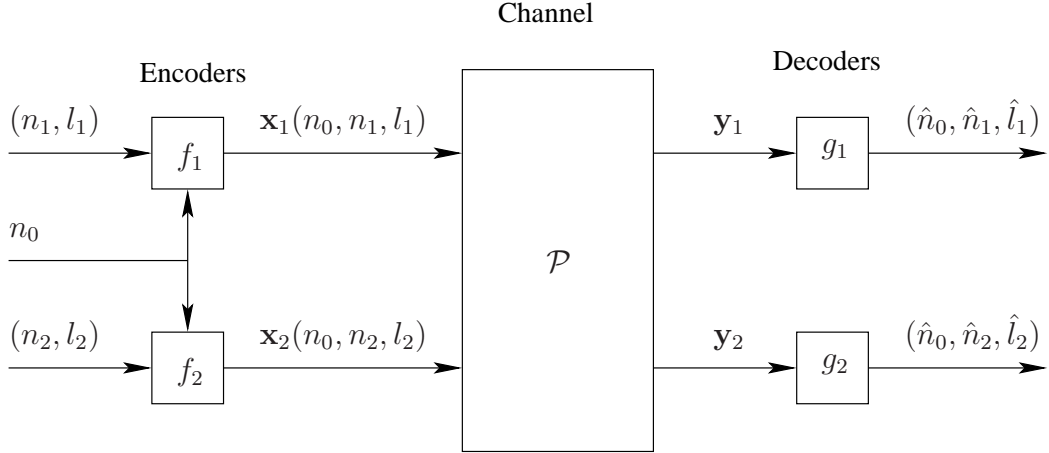

Fig. 2.   The modified interference channel with common information.

and two decoding functions

$$g_1 : \mathcal{Y}_1^n \to \mathcal{N}_0 \times \mathcal{N}_1 \times \mathcal{L}_1, \qquad\qquad g_2 : \mathcal{Y}_2^n \to \mathcal{N}_0 \times \mathcal{N}_2 \times \mathcal{L}_2,$$

such that $\max\{P_{e,1}^{(n)}, P_{e,2}^{(n)}\} \leq P_e$, where the average probabilities of decoding error denoted by $P_{e,1}^{(n)}$ and $P_{e,2}^{(n)}$ are computed as

$$P_{e,1}^{(n)} = \frac{1}{N_0 N_1 L_1 N_2 L_2} \sum_{n_0 n_1 l_1 n_2 l_2} p((\hat{n}_0, \hat{n}_1, \hat{l}_1) \neq (n_0, n_1, l_1) | (n_0, n_1, l_1, n_2, l_2)),$$

$$P_{e,2}^{(n)} = \frac{1}{N_0 N_1 L_1 N_2 L_2} \sum_{n_0 n_1 l_1 n_2 l_2} p((\hat{n}_0, \hat{n}_2, \hat{l}_2) \neq (n_0, n_2, l_2) | (n_0, n_1, l_1, n_2, l_2)).$$

A non-negative rate quintuple $(R_0, R_{12}, R_{11}, R_{21}, R_{22})$ is achievable for the channel $C_m$ if for any given $0 < P_e < 1$ and any sufficiently large $n$, there exists a $(2^{nR_0}, 2^{nR_{12}}, 2^{nR_{11}}, 2^{nR_{21}}, 2^{nR_{22}}, n, P_e)$ code for $C_m$.

*Remark 1:* It should be noted that compared with Fig. 2 in [9], our modified channel depicted in Fig. 2 does not include the index $\hat{n}_2$ (or $\hat{n}_1$) in the decoded message vector at decoder 1 (or decoder 2). This is due to the observation made in [15] that, although receiver 1 (or receiver 2) attempts to decode the crossly observable private message $n_2$ (or $n_1$), it is not essential to include decoding errors of such information in calculating probability of error at the respective receiver. This is also the reason why we call the two associated channels of an ICC as MACC-like channels instead of MACCs.

*Lemma 1:* If $(R_0, R_{12}, R_{11}, R_{21}, R_{22})$ is achievable for the channel $C_m$, then $(R_0, R_{12}+R_{11}, R_{21}+R_{22})$ is achievable for the associated ICC $C$.

*Remark 2:* Note that with the aid of Lemma 1, an achievable rate region for the modified ICC can be easily extended to one for the associated ICC.

## III. General Discrete Memoryless Interference channel with Common Information

### A. An Achievable Rate Region for the General Discrete Memoryless ICC

We first introduce three auxiliary random variables $U_0$, $U_1$ and $U_2$ that are defined over arbitrary finite sets $\mathcal{U}_0$, $\mathcal{U}_1$, and $\mathcal{U}_2$, respectively. Denote by $\mathcal{P}^*$ the set of all joint probability distributions $p(\cdot)$ that factor as

$$p(u_0, u_1, u_2, x_1, x_2, y_1, y_2) = p(u_0)p(u_1|u_0)p(u_2|u_0)p(x_1|u_1, u_0)p(x_2|u_2, u_0)p(y_1, y_2|x_1, x_2). \quad (1)$$

Let $\mathcal{R}_m(p)$ denote the set of all non-negative rate quintuples $(R_0, R_{12}, R_{11}, R_{21}, R_{22})$ such that

$$R_{11} \leq I(X_1; Y_1|U_0U_1U_2), \quad (2)$$

$$R_{12} + R_{11} \leq I(U_1X_1; Y_1|U_0U_2), \quad (3)$$

$$R_{11} + R_{21} \leq I(X_1U_2; Y_1|U_0U_1), \quad (4)$$

$$R_{12} + R_{11} + R_{21} \leq I(U_1X_1U_2; Y_1|U_0), \quad (5)$$

$$R_0 + R_{12} + R_{11} + R_{21} \leq I(U_0U_1X_1U_2; Y_1); \quad (6)$$

$$R_{22} \leq I(X_2; Y_2|U_0U_2U_1), \quad (7)$$

$$R_{21} + R_{22} \leq I(U_2X_2; Y_2|U_0U_1), \quad (8)$$

$$R_{22} + R_{12} \leq I(X_2U_1; Y_2|U_0U_2), \quad (9)$$

$$R_{21} + R_{22} + R_{12} \leq I(U_2X_2U_1; Y_2|U_0), \quad (10)$$

$$R_0 + R_{21} + R_{22} + R_{12} \leq I(U_0U_2X_2U_1; Y_2), \quad (11)$$

for some fixed joint probability distribution $p(\cdot) \in \mathcal{P}^*$. Note that each of the mutual information terms is computed with respect to the given fixed joint distribution.

*Theorem 1:* Any element $(R_0, R_{12}, R_{11}, R_{21}, R_{22}) \in \mathcal{R}_m(p)$ is achievable for the modified ICC $C_m$ for a fixed joint probability distribution $p(\cdot) \in \mathcal{P}^*$.

*Remark 3:* The lengthy proof is relegated to the last subsection of this section. Theorem 1 lays a foundation for us to establish an achievable rate region for the general ICC. One can interpret this achievable rate region as an intersection between the achievable rate regions of the two associated MACC-like channels, i.e., inequalities (2)–(6) depict the achievable rate region for one MACC-like channel, and inequalities (7)–(11) depict the other.

*Theorem 2:* The rate region $\mathcal{R}_m$ is achievable for the channel $C_m$ with $\mathcal{R}_m = \bigcup_{p(\cdot) \in \mathcal{P}^*} \mathcal{R}_m(p)$.

*Remark 4:* Theorem 2 is a direct extension of Theorem 1, and the proof is straightforward and omitted. Note that the rate region $\mathcal{R}_m$ is convex, and therefore no convex hull operation or time sharing is necessary. The proof of the convexity is given in the appendix.

Let us fix a joint distribution $p(\cdot) \in \mathcal{P}^*$, and denote by $\mathcal{R}(p)$ the set of all the non-negative rate triples $(R_0, R_1, R_2)$ such that $R_1 = R_{12} + R_{11}$ and $R_2 = R_{21} + R_{22}$ for some $(R_0, R_{12}, R_{11}, R_{21}, R_{22}) \in \mathcal{R}_m(p)$.

*Theorem 3:* $\mathcal{R}$ is an achievable rate region for the channel $C$ with $\mathcal{R} = \bigcup_{p(\cdot) \in \mathcal{P}^*} \mathcal{R}(p)$.

*Proof:* It suffices to prove that $\mathcal{R}(p)$ is an achievable rate region for $C$ for any fixed joint probability distribution $p(\cdot) \in \mathcal{P}^*$, while the achievability of any rate triple $(R_0, R_1, R_2) \in \mathcal{R}(p)$ follows immediately from Lemma 1 and Theorem 1. ∎

*Remark 5:* The main idea, as mentioned before, is that we allow the common information (of rate $R_0$) to be cooperatively transmitted by the two senders, on top of which we treat the private information at each sender as two parts. One part (of rate $R_{12}$ or $R_{21}$) of the private information at each sender is crossly observable to the non-pairing receiver, but not the other part (of rate $R_{11}$ or $R_{22}$). However, as discussed earlier, for each of the two receivers, the crossly observed information is not required to be decoded correctly. This will be elaborated more clearly in the proof of Theorem 1.

*Remark 6:* One can observe that the rate of the common information, $R_0$, is bounded by only one inequality at each decoder. This is similar to the case of MACC [19], [20], where the rate of the common information is bounded by only one inequality as well. This is due to the perfect cooperation of the two senders in transmitting the common information, and the simultaneous decoding. Details will also be illustrated in the proof of Theorem 1.

*Remark 7:* Our achievable rate region for the ICC is possibly a tight one, as we will demonstrate in Section IV that our region includes two well-known results as special cases. Moreover, in Section V we will show that our achievable rate region meets the outer bound for the capacity region of a class of DICCs, which results in the exact capacity region for this class of channels.

*Remark 8:* Note that the region $\mathcal{R}$ is also convex, and one can readily prove it by following procedures in the proof of the convexity of $\mathcal{R}_m$.

## B. An Explicit Description of the Achievable Rate Region

In order to reveal the geometric shape of the region of $\mathcal{R}$ depicted in Theorem 3, we derive an explicit description of the region by applying Fourier-Motzkin eliminations [9], [15], [21].

*Theorem 4:* The rate region $\mathcal{R}$ is achievable for the channel $C$ with $\mathcal{R} = \bigcup_{p(\cdot) \in \mathcal{P}^*} \mathcal{R}(p)$, where $\mathcal{R}(p)$ denotes the set of all rate triples $(R_0, R_1, R_2)$ such that

$$R_0 \leq I(U_0 U_1 X_1 U_2; Y_1),$$
$$R_0 \leq I(U_0 U_2 X_2 U_1; Y_2),$$
$$R_1 \leq I(U_1 X_1; Y_1 | U_0 U_2),$$
$$R_2 \leq I(U_2 X_2; Y_2 | U_0 U_1),$$
$$R_1 + R_2 \leq I(X_1 U_2; Y_1 | U_0 U_1) + I(X_2 U_1; Y_2 | U_0 U_2);$$
$$R_1 + R_2 \leq I(U_1 X_1 U_2; Y_1 | U_0) + I(X_2; Y_2 | U_0 U_1 U_2),$$
$$R_0 + R_1 + R_2 \leq I(U_0 U_1 X_1 U_2; Y_1) + I(X_2; Y_2 | U_0 U_1 U_2);$$
$$R_1 + R_2 \leq I(X_1; Y_1 | U_0 U_1 U_2) + I(U_2 X_2 U_1; Y_2 | U_0),$$
$$R_0 + R_1 + R_2 \leq I(X_1; Y_1 | U_0 U_1 U_2) + I(U_0 U_2 X_2 U_1; Y_2);$$
$$2R_1 + R_2 \leq I(U_1 X_1 U_2; Y_1 | U_0) + I(X_1; Y_1 | U_0 U_1 U_2) + I(X_2 U_1; Y_2 | U_0 U_2),$$
$$R_0 + 2R_1 + R_2 \leq I(U_0 U_1 X_1 U_2; Y_1) + I(X_1; Y_1 | U_0 U_1 U_2) + I(X_2 U_1; Y_2 | U_0 U_2);$$
$$R_1 + 2R_2 \leq I(U_2 X_2 U_1; Y_2 | U_0) + I(X_2; Y_2 | U_0 U_1 U_2) + I(X_1 U_2; Y_1 | U_0 U_1),$$
$$R_0 + R_1 + 2R_2 \leq I(U_0 U_2 X_2 U_1; Y_2) + I(X_2; Y_2 | U_0 U_1 U_2) + I(X_1 U_2; Y_1 | U_0 U_1),$$

for some fixed joint distribution $p(\cdot) \in \mathcal{P}^*$.

*Remark 9:* The close relation between the explicit Chong-Motani-Garg region and the capacity region of a class of deterministic ICs given in [11] was pointed out in [21]. Similarly, we will disclose that the explicit region for the ICC is also closely related to the capacity region of a class of DICCs investigated in Section V.

### C. The Proof of Theorem 1

In this section, we will prove Theorem 1 that is the core of this paper up to here. The general idea is to apply the cascaded superposition encoding and simultaneous decoding. As the following lemma will be frequently used, we state it here before the proof of Theorem 1.

*Lemma 2 ( [22, Theorem 14.2.3]):* Let $A_\epsilon^{(n)}$ denote the typical set for the probability distribution $p(s_1, s_2, s_3)$, and let

$$P(\mathbf{S}_1' = \mathbf{s}_1, \mathbf{S}_2' = \mathbf{s}_2, \mathbf{S}_3' = \mathbf{s}_3) = \prod_{i=1}^n p(s_{1i}|s_{3i})p(s_{2i}|s_{3i})p(s_{3i}), \qquad (12)$$

then

$$P\{(\mathbf{S}_1', \mathbf{S}_2', \mathbf{S}_3') \in A_\epsilon^{(n)}\} \doteq 2^{-n(I(S_1;S_2|S_3)\pm 6\epsilon)}. \qquad (13)$$

*Proof of Theorem 1:* [**Codebook Generation**.] Let us fix a joint distribution $p(\cdot)$ that factors in the form of (1). We first generate $2^{nR_0}$ independent codewords $\mathbf{u}_0(i)$, $i \in \{1, ..., 2^{nR_0}\}$, according to $\prod_{i=1}^n p(u_{0i})$. At encoder 1, for each codeword $\mathbf{u}_0(i)$, generate $2^{nR_{12}}$ independent codewords $\mathbf{u}_1(i, j)$, $j \in \{1, ... 2^{nR_{12}}\}$ according to $\prod_{i=1}^n p(u_{1i}|u_{0i})$. Subsequently, for each pair of codewords $(\mathbf{u}_0(i), \mathbf{u}_1(i, j))$, generate $2^{nR_{11}}$ independent codewords $\mathbf{x}_1(i, j, k)$, $k \in \{1, ... 2^{nR_{11}}\}$, according to $\prod_{i=1}^n p(x_{1i}|u_{1i}u_{0i})$. Similarly at encoder 2, for each codeword $\mathbf{u}_0(i)$, generate $2^{nR_{21}}$ independent codewords $\mathbf{u}_2(i, l)$, $l \in \{1, ... 2^{nR_{21}}\}$ according to $\prod_{i=1}^n p(u_{2i}|u_{0i})$. Subsequently, for each codeword pair $(\mathbf{u}_0(i), \mathbf{u}_2(i, l))$, generate $2^{nR_{22}}$ independent codewords $\mathbf{x}_2(i, l, m)$, $m \in \{1, ... 2^{nR_{22}}\}$, according to $\prod_{i=1}^n p(x_{2i}|u_{2i}u_{0i})$. The entire codebook consisting of all the codewords $\mathbf{u}_0(i)$, $\mathbf{u}_1(i, j)$, $\mathbf{x}_1(i, j, k)$, $\mathbf{u}_2(i, l)$ and $\mathbf{x}_2(i, l, m)$ with $i \in \{1, ..., 2^{nR_0}\}$, $j \in \{1, ..., 2^{nR_{12}}\}$, $k \in \{1, ..., 2^{nR_{11}}\}$, $l \in \{1, ..., 2^{nR_{21}}\}$ and $m \in \{1, ..., 2^{nR_{22}}\}$ is revealed to both receivers.

[**Encoding** & **Transmission**.] Suppose that the source message vector generated at the two senders is $(n_0, n_1, l_1, n_2, l_2) = (i, j, k, l, m)$. Sender 1 transmits codeword $\mathbf{x}_1(i, j, k)$ with $n$ channel uses, while sender 2 transmits codeword $\mathbf{x}_2(i, l, m)$ with $n$ channel uses. The transmissions are assumed to be perfectly synchronized.

[**Decoding**.] Each of the receivers accumulates an $n$-length channel output sequence, $\mathbf{y}_1$ (receiver 1) or $\mathbf{y}_2$ (receiver 2). Let $A_\epsilon^{(n)}$ denote the typical sets of the respective joint distributions. Decoder 1 declares that $(\hat{i}, \hat{j}, \hat{k})$ is received, if $(\hat{i}, \hat{j}, \hat{k})$ is the unique message vector such that $(\mathbf{u}_0(\hat{i}), \mathbf{u}_1(\hat{i}, \hat{j}), \mathbf{x}_1(\hat{i}, \hat{j}, \hat{k}), \mathbf{u}_2(\hat{i}, l), \mathbf{y}_1) \in A_\epsilon^{(n)}$ for some $l$; otherwise, a decoding error is declared. Similarly, decoder 2 looks for a unique message vector $(\hat{i}, \hat{l}, \hat{m})$ such that $(\mathbf{u}_0(\hat{i}), \mathbf{u}_2(\hat{i}, \hat{l}), \mathbf{x}_2(\hat{i}, \hat{l}, \hat{m}), \mathbf{u}_1(\hat{i}, j), \mathbf{y}_2) \in A_\epsilon^{(n)}$ for some $j$; otherwise, a decoding error is declared.

[**Analysis of the Probability of Decoding Error**.] Because of the symmetry of the codebook generation, the probability of error does not depend on which message vector is encoded and transmitted. Since the messages are uniformly generated over their respective ranges, the average error probability over all the possible messages is equal to the probability of error incurred when any message vector is encoded and transmitted. We hence only analyze the probability of decoding error for decoder 1 in details, since the same analysis can be carried out for decoder 2. Without loss of generality, we assume that a source message vector $(n_0, n_l, l_1, n_2, l_2) = (1, 1, 1, 1, 1)$ is encoded and transmitted for the subsequent analysis. We first define the event

$$E_{ijkl} \triangleq \{(\mathbf{U}_0(i), \mathbf{U}_1(i,j), \mathbf{X}_1(i,j,k), \mathbf{U}_2(i,l), \mathbf{Y}_1) \in A_\epsilon^{(n)}\}.$$

The possible error events can be grouped into two classes: 1) the codewords transmitted are not jointly typical, i.e., $E_{1111}^c$ happens; 2) there exist some $(i,j,k) \neq (1,1,1)$ such that $E_{ijkl}$ happens ($l$ may not be 1). Thus the probability of decoding error at decoder 1 can be expressed as

$$P_{e,1}^{(n)} = P(E_{1111}^c \bigcup \cup_{(i,j,k) \neq (1,1,1)} E_{ijkl}). \tag{14}$$

By applying the union bound, we can upper-bound (14) as

$$
\begin{aligned}
P_{e,1}^{(n)} \leq &P(E_{1111}^c) + P(\cup_{(i,j,k) \neq (1,1,1)} E_{ijkl}) \\
\leq &P(E_{1111}^c) + \sum_{i \neq 1} P(E_{i111}) + \sum_{i \neq 1, l \neq 1} P(E_{i11l}) + \sum_{j \neq 1} P(E_{1j11}) + \sum_{j \neq 1, l \neq 1} P(E_{1j1l}) \\
&+ \sum_{k \neq 1} P(E_{11k1}) + \sum_{k \neq 1, l \neq 1} P(E_{11kl}) + \sum_{i \neq 1, j \neq 1} P(E_{ij11}) + \sum_{i \neq 1, j \neq 1, l \neq 1} P(E_{ij1l}) \\
&+ \sum_{i \neq 1, k \neq 1} P(E_{i1k1}) + \sum_{i \neq 1, k \neq 1, l \neq 1} P(E_{i1kl}) + \sum_{j \neq 1, k \neq 1} P(E_{1jk1}) \\
&+ \sum_{j \neq 1, k \neq 1, l \neq 1} P(E_{1jkl}) + \sum_{i \neq 1, j \neq 1, k \neq 1} P(E_{ijk1}) + \sum_{i \neq 1, j \neq 1, k \neq 1, l \neq 1} P(E_{ijkl}). \tag{15}
\end{aligned}
$$

Due to the asymptotic equipartition property (AEP) [22], $P(E_{1111}^c)$ in (15) can be made arbitrarily small as long as $n$ is sufficiently large. The rest of the fourteen probability terms in (15) can be evaluated through one standard procedure, which is demonstrated as follows. To evaluate $P(E_{i111})$, we apply Lemma 2 by letting $\mathbf{S}_1' = (\mathbf{U}_0(i), \mathbf{U}_1(i,1), \mathbf{X}_1(i,1,1), \mathbf{U}_2(i,1))$, $\mathbf{S}_2' = \mathbf{Y}_1$, and $\mathbf{S}_3' = \emptyset$, where $\emptyset$ denotes the empty set. Note that the assumption of the lemma on the joint distribution of $\mathbf{S}_1'$, $\mathbf{S}_2'$ and $\mathbf{S}_3'$ is satisfied, and thus it follows that

$$P(E_{i111}) \leq 2^{-n(I(U_0 U_1 X_1 U_2; Y_1) - 6\epsilon)}. \tag{16}$$

Since the case with $\mathbf{S}_3' = \emptyset$ seems not archetypal, we evaluate one more probability term, $P(E_{1jk1})$. Again, we use Lemma 2 by letting $\mathbf{S}_1' = (\mathbf{U}_1(1,j), \mathbf{X}_1(1,j,k))$, $\mathbf{S}_2' = \mathbf{Y}_1$, and $\mathbf{S}_3' = (\mathbf{U}_0(1), \mathbf{U}_2(1,1))$ to obtain

$$P(E_{1jk1}) \leq 2^{-n(I(U_1 X_1; Y_1 | U_0 U_2) - 6\epsilon)}. \tag{17}$$

By repeatedly applying Lemma 2, we obtain upper-bounds of the remaining twelve probability terms. Further, we employ these bounds to derive an upper-bound of the probability of decoding

error at decoder 1 as

$$
\begin{aligned}
P_{e,1}^{(n)} \leq \;& \epsilon + 2^{nR_0} 2^{-n(I(U_0 U_1 X_1 U_2; Y_1) - 6\epsilon)} + 2^{n(R_0 + R_{21})} 2^{-n(I(U_0 U_1 X_1 U_2; Y_1) - 6\epsilon)} \\
& + 2^{nR_{12}} 2^{-n(I(U_1 X_1; Y_1 | U_0 U_2) - 6\epsilon)} + 2^{n(R_{12} + R_{21})} 2^{-n(I(U_1 X_1 U_2; Y_1 | U_0) - 6\epsilon)} \\
& + 2^{nR_{11}} 2^{-n(I(X_1; Y_1 | U_0 U_1 U_2) - 6\epsilon)} + 2^{n(R_{11} + R_{21})} 2^{-n(I(X_1 U_2; Y_1 | U_0 U_1) - 6\epsilon)} \\
& + 2^{n(R_0 + R_{12})} 2^{-n(I(U_0 U_1 X_1 U_2; Y_1) - 6\epsilon)} + 2^{n(R_0 + R_{12} + R_{21})} 2^{-n(I(U_0 U_1 X_1 U_2; Y_1) - 6\epsilon)} \\
& + 2^{n(R_0 + R_{11})} 2^{-n(I(U_0 U_1 X_1 U_2; Y_1) - 6\epsilon)} + 2^{n(R_0 + R_{11} + R_{21})} 2^{-n(I(U_0 U_1 X_1 U_2; Y_1) - 6\epsilon)} \\
& + 2^{n(R_{12} + R_{11})} 2^{-n(I(U_1 X_1; Y_1 | U_0 U_2) - 6\epsilon)} + 2^{n(R_{12} + R_{11} + R_{21})} 2^{-n(I(U_1 X_1 U_2; Y_1 | U_0) - 6\epsilon)} \\
& + 2^{n(R_0 + R_{12} + R_{11})} 2^{-n(I(U_0 U_1 X_1 U_2; Y_1) - 6\epsilon)} + 2^{n(R_0 + R_{12} + R_{11} + R_{21})} 2^{-n(I(U_0 U_1 X_1 U_2; Y_1) - 6\epsilon)}. \quad (18)
\end{aligned}
$$

It is now easy to check that when inequalities (2)–(6) hold and $n$ is sufficiently large, we have

$$
P_{e,1}^{(n)} \leq 15\epsilon. \tag{19}
$$

By symmetry, the decoding error probability becomes $P_{e,2}^{(n)} \leq 15\epsilon$ for decoder 2, when inequalities (7)–(11) hold and $n$ is sufficiently large. It follows that $\max\{P_{e,1}^{(n)}, P_{e,2}^{(n)}\} \leq 15\epsilon$, and thus any rate quintuple $(R_0, R_{12}, R_{11}, R_{21}, R_{22}) \in \mathcal{R}_m(p)$ is achievable for the modified ICC $C_m$ for a fixed joint distribution $p(\cdot) \in \mathcal{P}^*$. ∎

*Remark 10:* In what follows, we list a few remarks on the encoding and decoding scheme used in our derivation.

1) We term the above coding scheme "the cascaded superposition coding", because there are three layers of code with the bottom one $\mathbf{u}_0(i)$ carrying the common information. The second layer consists of $\mathbf{u}_1(i, j)$ and $\mathbf{u}_2(i, l)$. This layer superimposes the part of each sender's private information, which is crossly observable to the non-pairing receiver, on the bottom layer; while $\mathbf{x}_1(i, j, k)$ and $\mathbf{x}_2(i, l, m)$ form the top layer, and they are generated by superimposing the part of private information which is not crossly observable on top of both the second layer and the bottom layer.

2) The encoding scheme is auxiliary random variable efficient in the sense that it only requires *three* auxiliary random variables instead of *five* required if one follows [9] to apply the simultaneous superposition coding scheme. It not only greatly simplifies the description of the achievable rate region in terms of the number of inequalities required, but also has implications on practical code design or implementation of the system in the sense that the number of different codes required is reduced.

3) For the decoding, the simultaneous joint typicality of three layers of codes is examined. It is the reason why we could have to use fourteen inequalities due to (18), but we in fact only use five inequalities (inequalities (2)–(6)) instead. Due to the cascaded superpositioning and simultaneous decoding, $R_0$ is only bounded together with other rates by (6) or (11) for each decoder. The advantage of the simultaneous decoding over the successive decoding is also demonstrated with an example of MACC in [23].

## IV. SOME SPECIAL CASES OF THE ICCs

### A. Strong Interference Channel with Common Information

We demonstrate that the capacity region of the SICC given in [18] can be obtained as a special case of our achievable rate region for the general ICC.

Let $\mathcal{P}_s$ denote the set of all joint distributions $p(u_0, x_1, x_2, y_1, y_2)$ that factor as $p(u_0)p(x_1|u_0)$ $p(x_2|u_0)p(y_1, y_2|x_1, x_2)$. As defined in [18], an ICC is considered as a SICC if

$$I(X_1; Y_1|X_2U_0) \leq I(X_1; Y_2|X_2U_0), \tag{20}$$

$$I(X_2; Y_2|X_1U_0) \leq I(X_2; Y_1|X_1U_0), \tag{21}$$

for all joint probability distributions $p(\cdot) \in \mathcal{P}_s$. Let $\mathcal{R}_s(p)$ denote the set of all non-negative rate triples $(R_0, R_1, R_2)$ such that

$$R_1 \leq I(X_1; Y_1|X_2U_0), \tag{22}$$

$$R_2 \leq I(X_2; Y_2|X_1U_0), \tag{23}$$

$$R_1 + R_2 \leq \min\{I(X_1X_2; Y_1|U_0), I(X_2X_1; Y_2|U_0)\}, \tag{24}$$

$$R_0 + R_1 + R_2 \leq \min\{I(X_1X_2; Y_1), I(X_2X_1; Y_2)\}, \tag{25}$$

for a fixed joint distribution $p(\cdot) \in \mathcal{P}_s$.

*Corollary 1 ( [18, Achievability of Theorem 1]):* Any rate triple $(R_0, R_1, R_2) \in \mathcal{C}_s$ is achievable for the SICC with $\mathcal{C}_s = \bigcup_{p(\cdot) \in \mathcal{P}_s} \mathcal{R}_s(p)$.

*Proof:* It suffices to show that $\mathcal{R}_s(p)$ is achievable for a fixed joint distribution $p(\cdot) \in \mathcal{P}_s$. Referring to the region defined by (2)–(11), we set $U_1 = X_1$ and $U_2 = X_2$, which makes both $R_{11}$ and $R_{22}$ become zero; and we substitute $R_{12}$ with $R_1$, and $R_{21}$ with $R_2$. Hence, inequalities (2)–(11) reduce to

$$R_1 \leq I(X_1; Y_1|U_0X_2), \tag{26}$$

$$R_2 \leq I(X_2; Y_1|U_0X_1), \tag{27}$$

$$R_1 + R_2 \leq I(X_1X_1; Y_1|U_0), \tag{28}$$

$$R_0 + R_1 + R_2 \leq I(U_0X_1X_2; Y_1); \tag{29}$$

$$R_2 \leq I(X_2; Y_2|U_0X_1), \tag{30}$$

$$R_1 \leq I(X_1; Y_2|U_0X_2), \tag{31}$$

$$R_2 + R_1 \leq I(X_2X_1; Y_2|U_0), \tag{32}$$

$$R_0 + R_2 + R_1 \leq I(U_0X_2X_1; Y_2). \tag{33}$$

Since for the SICC inequality (20) must hold for the given joint distribution, inequality (26) implies (31), and thus inequality (31) can be excluded. Similarly, inequality (27) can be excluded as well. Due to the fact that $U_0$, $(X_1, X_2)$ and $Y_t$, $t = 1, 2$, form a Markov chain, $I(U_0X_1X_2; Y_1) = I(X_1X_2; Y_1)$ and $I(U_0X_1X_2; Y_2) = I(X_1X_2; Y_2)$. Hence, $\mathcal{R}_s(p)$ is an achievable rate region for the SICC for a fixed joint distribution $p(\cdot) \in \mathcal{P}_s$, and $\mathcal{C}_s$ is achievable for the SICC.  ■

*Remark 11:* By letting $U_1 = X_1$ and $U_2 = X_2$, we treat the private information at each sender as a whole instead of two parts. This differs from what was mentioned earlier in Remark 5. Here the full private information at each sender is allowed to be crossly observed by the respective non-pairing receivers due to the strong interference. In fact, inequalities (26)–(33) also define one achievable rate region for the general ICC. However, it is only tight for the case of strong interference, but not for the general case.

## B. Interference Channel without Common Information

We now consider the general IC (without common information) as a special case of the ICC, and demonstrate that our achievable rate region for the ICC subsumes the Chong-Motani-Garg region [15] as a special case. Note that the Chong-Motani-Garg region is one of the two best achievable rate regions for the IC (without common information), and it has a much simpler description of the region compared to the other.

Let $Q$ denote a time sharing random variable defined over an arbitrary finite alphabet $\mathcal{Q}$, and $\mathcal{P}_o$ denote the set of all joint distributions that factor as

$$p(q, u_1, u_2, x_1, x_2, y_1, y_2) = p(q)p(u_1|q)p(u_2|q)p(x_1|u_1, q)p(x_2|u_2, q)p(y_1, y_2|x_1, x_2). \quad (34)$$

Let $\mathcal{R}_o(p)$ denote the set of all rate pairs $(R_1, R_2)$ with $R_1 = R_{12} + R_{11}$ and $R_2 = R_{21} + R_{22}$ such that

$$R_{11} \leq I(X_1; Y_1|U_1U_2Q), \quad (35)$$

$$R_{12} + R_{11} \leq I(X_1; Y_1|U_2Q), \quad (36)$$

$$R_{11} + R_{21} \leq I(X_1U_2; Y_1|U_1Q), \quad (37)$$

$$R_{12} + R_{11} + R_{21} \leq I(X_1U_2; Y_1|Q); \quad (38)$$

$$R_{22} \leq I(X_2; Y_2|U_2U_1Q), \quad (39)$$

$$R_{21} + R_{22} \leq I(X_2; Y_2|U_1Q), \quad (40)$$

$$R_{22} + R_{12} \leq I(X_2U_1; Y_2|U_2Q), \quad (41)$$

$$R_{21} + R_{22} + R_{12} \leq I(X_2U_1; Y_2|Q), \quad (42)$$

for a fixed joint distribution $p(\cdot) \in \mathcal{P}_o$.

*Corollary 2 ( [15, Theorem 3]):* $\mathcal{R}_o$ is an achievable rate region for the IC with

$$\mathcal{R}_o = \bigcup_{p(\cdot) \in \mathcal{P}_o} \mathcal{R}_o(p).$$

*Proof:* It suffices to show that $\mathcal{R}_o(p)$ is achievable for a fix joint distribution $p(\cdot) \in \mathcal{P}_o$. We still work on the region defined by (2)–(11) with respect to the general ICC. Since there is no common information, we set $\mathcal{U}_0 = \emptyset$, and $R_0 = 0$. Note that the existence of $U_0$ in fact contributes to the convexity of the rate region $\mathcal{R}_m$, which one can observe from the proof of the convexity of $\mathcal{R}_m$ in the appendix. When $U_0$ is dropped, we need introduce the time sharing random variable $Q$ to maintain the convexity. The rate region defined by (2)–(11) now becomes

$$R_{11} \leq I(X_1; Y_1|U_1U_2Q), \quad (43)$$

$$R_{12} + R_{11} \leq I(U_1X_1; Y_1|U_2Q), \quad (44)$$

$$R_{11} + R_{21} \leq I(X_1U_2; Y_1|U_1Q), \quad (45)$$

$$R_{12} + R_{11} + R_{21} \leq I(U_1X_1U_2; Y_1|Q); \quad (46)$$

$$R_{22} \leq I(X_2; Y_2|U_2U_1Q), \quad (47)$$

$$R_{21} + R_{22} \leq I(U_2X_2; Y_2|U_1Q), \quad (48)$$

$$R_{22} + R_{12} \leq I(X_2U_1; Y_2|U_2Q), \quad (49)$$

$$R_{21} + R_{22} + R_{12} \leq I(U_2X_2U_1; Y_2|Q). \quad (50)$$

According to the joint probability distribution $p(\cdot) \in \mathcal{P}_o$, the random variables $U_1$, $X_1$ and $Y_1$ form a Markov chain conditioned on $U_2$ and $Q$, and thus $I(U_1 X_1; Y_1 | U_2 Q) = I(X_1; Y_1 | U_2 Q)$ in (44). Similar simplifications can be made on (46), (48) and (50). Finally, the derived region becomes the same as one described by (35)–(42). With Lemma 1, we conclude that $\mathcal{R}_o(p)$ is achievable for a fix joint distribution $p(\cdot) \in \mathcal{P}_o$, and $\mathcal{R}_o$ is achievable for the IC (without common information). ∎

*Remark 12:* As shown above, our achievable rate region for the ICC in the implicit form subsumes the implicit Chong-Motani-Garg region as a special case. Alternatively, following the same procedures as demonstrated in the above proof, we can work on the explicit region given in Theorem 4 to obtain the explicit Chong-Motani-Garg region [15, Theorem 4] as well.

### C. Asymmetric Interference Channel with Common Information

In this subsection, we investigate a class of the ICCs where one of the two senders does not have private information to transmit, and we term this class of channels as the asymmetric interference channel with common information (AICC). We present an achievable rate region for the AICC as a byproduct of our result for the general ICC.

Without loss of generality, we assume that sender 1 only has the common message $w_0$ to send to receiver 1, while sender 2 needs transmit both the common message $w_0$ and the private message $w_2$ to receiver 2. Fig. 3 depicts an AICC, which we denote by $C_a$. Following the definitions and channel models given in Section II, we can easily obtain a corresponding modified channel as shown in Fig. 4 for $C_a$, and we denote it by $C_a^m$. Note that the capacity region of $C_a^m$ is a set of all achievable rate triples $(R_0, R_{21}, R_{22})$, whereas the capacity region of $C_a$ is a set of all achievable rate pairs $(R_0, R_2)$.

PSfrag replacements

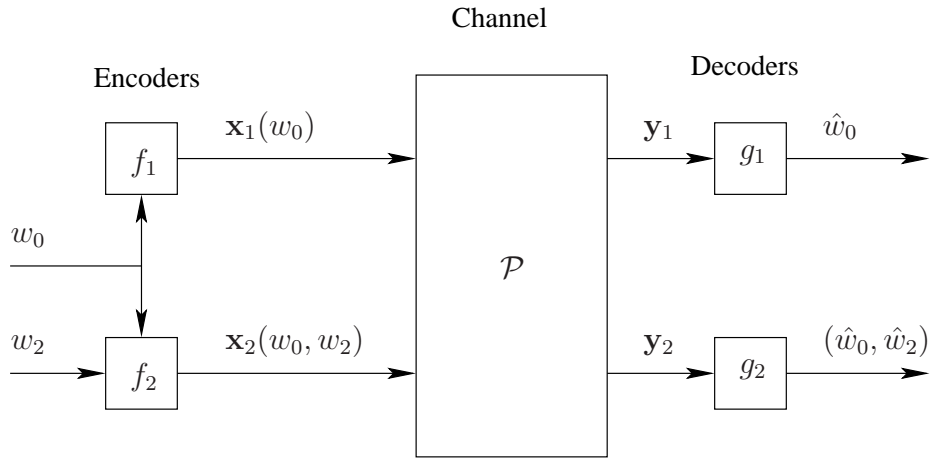

Fig. 3. The asymmetric interference channel with common information.

Let $\mathcal{P}_a$ denote the set of all joint distributions

$$p(x_1, u_2, x_2, y_1, y_2) = p(x_1, u_2, x_2) p(y_1, y_2 | x_1, x_2), \qquad (51)$$

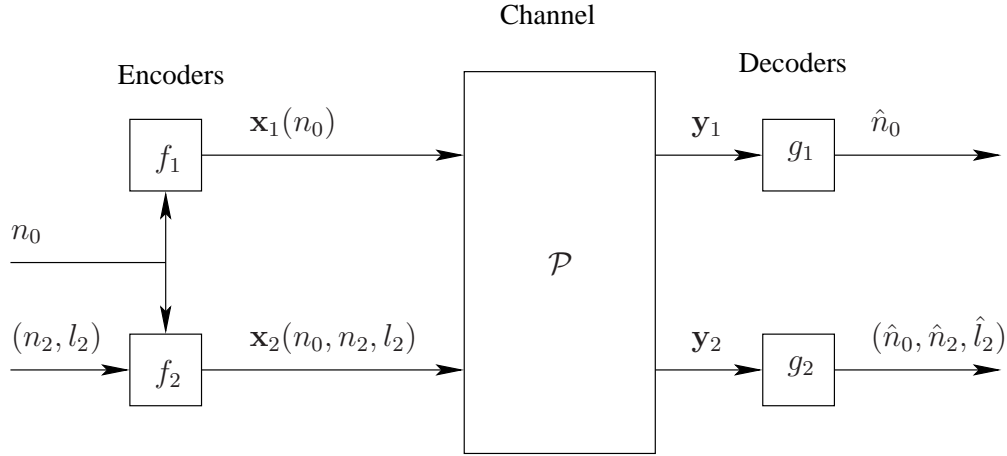

Fig. 4. The modified asymmetric interference channel with common information.

and let $\mathcal{R}_a^m(p)$ denote the set of all rate triples $(R_0, R_{21}, R_{22})$ such that

$$R_0 + R_{21} \leq I(X_1 U_2; Y_1), \tag{52}$$

$$R_{22} \leq I(X_2; Y_2 | U_2 X_1), \tag{53}$$

$$R_{21} + R_{22} \leq I(X_2; Y_2 | X_1), \tag{54}$$

$$R_0 + R_{21} + R_{22} \leq I(X_1 X_2; Y_2), \tag{55}$$

for some fixed joint distribution $p(\cdot) \in \mathcal{P}_a$.

*Corollary 3:* $\mathcal{R}_a^m$ is an achievable rate region for the modified channel $C_a^m$ with $\mathcal{R}_a^m = \bigcup_{p(\cdot) \in \mathcal{P}_a} \mathcal{R}_a^m(p)$.

*Remark 13:* By setting $R_{12} = 0$ and $R_{11} = 0$, and substituting both $U_0$ and $U_1$ with $X_1$, one can easily obtain Corollary 3 from Theorem 2.

By applying Fourier-Motzkin elimination on (52)–(55) with $R_2 = R_{21} + R_{22}$, we obtain an explicit achievable rate region for the AICC as follows.

*Theorem 5:* $\mathcal{R}_a = \bigcup_{p(\cdot) \in \mathcal{P}_a} \mathcal{R}_a(p)$ is an achievable rate region for $C_a$, where $\mathcal{R}_a(p)$ is the set of all rate pairs $(R_0, R_2)$ such that

$$R_0 \leq I(X_1 U_2; Y_1),$$

$$R_2 \leq I(X_2; Y_2 | X_1),$$

$$R_0 + R_2 \leq \min\{I(X_1 X_2; Y_2), I(X_1 U_2; Y_1) + I(X_2; Y_2 | U_2 X_1)\},$$

for some fixed joint distribution $p(\cdot) \in \mathcal{P}_a$.

*Remark 14:* 1) Alternatively, one can obtain Theorem 5 from Theorem 4 by letting $R_1 = 0$ and substituting $U_0$ and $U_1$ with $X_1$. 2) The coding strategy for this channel remains generally the same as the one for the general ICC: both senders first need cooperate to transmit the common information; while sender 2 treats the private information as two parts with one part crossly observable to receiver 1 but not the other part. 3) Although there is only one auxiliary random variable involved, the converse remains extremely difficult to establish.

## V. THE CAPACITY REGION OF A CLASS OF DETERMINISTIC INTERFERENCE CHANNELS WITH COMMON INFORMATION

In this section, we investigate a class of discrete memoryless DICCs as depicted in Fig. 5. The main characteristics of the channel remain the same as those of an ICC, i.e., source messages $(w_0, w_1, w_2)$, channel input and output alphabets $\mathcal{X}_t$ and $\mathcal{Y}_t$, $t = 1, 2$, encoding functions ($f_1(\cdot)$ and $f_2(\cdot)$) and decoding functions ($g_1(\cdot)$ and $g_2(\cdot)$), existence of codes and achievable rates are defined the same as those for the general ICC. The distinction lies on the channel transition, which is governed by the following deterministic functions:

PSfrag replacements

$$V_t = k_t(X_t), \quad t = 1, 2; \tag{56}$$

$$Y_1 = o_1(X_1, V_2), \tag{57}$$

$$Y_2 = o_2(X_2, V_1), \tag{58}$$

where $V_1$ and $V_2$ represent the interference signals caused by $X_1$ and $X_2$ at the corresponding receivers. Furthermore, we assume that there exist two more deterministic functions, $V_2 = h_1(Y_1, X_1)$ and $V_1 = h_2(Y_2, X_2)$. We denote this class of DICCs by $C_d$.

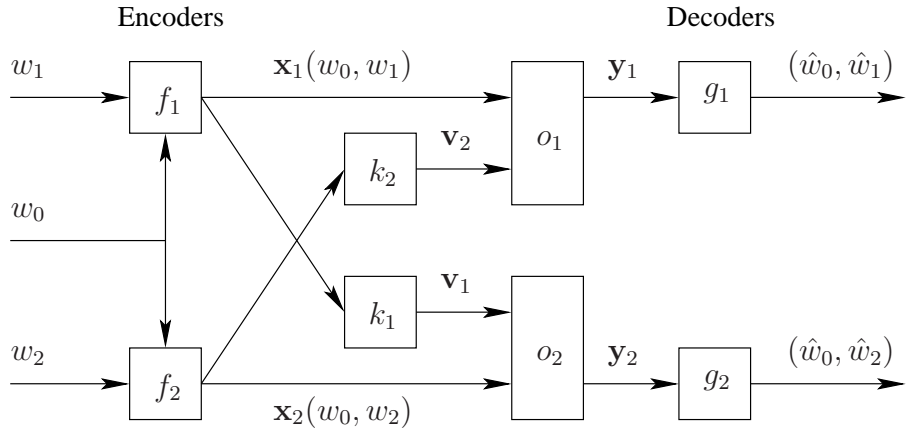

Fig. 5.   The class of deterministic interference channels with common information.

Note that the channel defined above is similar to the one investigated in [11], but there is a slight difference. In [11], it is required that $H(Y_1|X_1) = H(V_2)$ and $H(Y_2|X_2) = H(V_1)$ for all product distributions of $X_1 X_2$. It has also been pointed out in [11] that this requirement is equivalent to requiring the existence of $V_2 = h_1(Y_1, X_1)$ and $V_1 = h_2(Y_2, X_2)$. Nevertheless, we require the latter rather than the former, and in fact the former is not satisfied in our case. We will demonstrate that $V_2 = h_1(Y_1, X_1)$ and $V_1 = h_2(Y_2, X_2)$ are the actual governing conditions for this class of DICCs.

Let $\mathcal{P}_d$ denote the set of all joint distributions $p(\cdot)$ that factor as

$$p(v_0, x_1, x_2) = p(v_0)p(x_1|v_0)p(x_2|v_0), \tag{59}$$

where $v_0$ is the realization of an auxiliary random variable $V_0$ defined over an arbitrary finite set

$\mathcal{V}_0$. Let $\mathcal{R}_d(p)$ denote the set of all rate triples $(R_0, R_1, R_2)$ such that

$$R_0 \leq H(Y_1), \tag{60}$$

$$R_0 \leq H(Y_2), \tag{61}$$

$$R_1 \leq H(Y_1|V_0V_2), \tag{62}$$

$$R_2 \leq H(Y_2|V_0V_1), \tag{63}$$

$$R_1 + R_2 \leq H(Y_1|V_0V_1) + H(Y_2|V_0V_2); \tag{64}$$

$$R_1 + R_2 \leq H(Y_1|V_0) + H(Y_2|V_0V_1V_2), \tag{65}$$

$$R_0 + R_1 + R_2 \leq H(Y_1) + H(Y_2|V_0V_1V_2); \tag{66}$$

$$R_1 + R_2 \leq H(Y_1|V_0V_1V_2) + H(Y_2|V_0), \tag{67}$$

$$R_0 + R_1 + R_2 \leq H(Y_1|V_0V_1V_2) + H(Y_2); \tag{68}$$

$$2R_1 + R_2 \leq H(Y_1|V_0) + H(Y_1|V_0V_1V_2) + H(Y_2|V_0V_2), \tag{69}$$

$$R_0 + 2R_1 + R_2 \leq H(Y_1) + H(Y_1|V_0V_1V_2) + H(Y_2|V_0V_2); \tag{70}$$

$$R_1 + 2R_2 \leq H(Y_2|V_0) + H(Y_2|V_0V_1V_2) + H(Y_1|V_0V_1), \tag{71}$$

$$R_0 + R_1 + 2R_2 \leq H(Y_2) + H(Y_2|V_0V_1V_2) + H(Y_1|V_0V_1), \tag{72}$$

for some fixed joint distribution $p(\cdot) \in \mathcal{P}_d$.

*Theorem 6:* The capacity region of $C_d$ is the closure of $\bigcup_{p(\cdot) \in \mathcal{P}_d} \mathcal{R}_d(p)$.

*Proof:* 1) **Achievability**: It suffices to show that $\mathcal{R}_d(p)$ is achievable for the channel $C_d$ for a fixed joint distribution $p(\cdot) \in \mathcal{P}_d$. As the joint distribution $p(\cdot) \in \mathcal{P}_d$ does not involve $V_1$ and $V_2$, it appears incurring difficulty for us to apply the cascaded superposition coding strategy developed for the general ICC to this channel, due to the lack of auxiliary random variables. Nevertheless, because the interferences $V_1$ and $V_2$ are determined by the channel inputs $X_1$ and $X_2$, we can extend the joint distribution in the form of (59) to one containing $V_1$ and $V_2$ as

$$p(v_0, x_1, x_2, v_1, v_2) = p(v_0)p(x_1|v_0)p(x_2|v_0)\delta(v_1 - k_1(x_1))\delta(v_2 - k_1(x_2)), \tag{73}$$

where $\delta(\cdot)$ is the Kronecker Delta function. Since $X_1$ and $X_2$ are conditionally independent given $V_0$, the interferences $V_1$ and $V_2$ also become conditionally independent given $V_0$. Therefore, the extended joint distribution (73) can be factored as

$$p(v_0, x_1, x_2, v_1, v_2) = p(v_0)p(v_1|v_0)p(v_2|v_0)p(x_1|v_1, v_0)p(x_2|v_2, v_0),$$

and the achievability of the region $\mathcal{R}_d(p)$ follows readily from Theorem 4.

2) **Converse**: It suffices to show that for any $(2^{nR_0}, 2^{nR_1}, 2^{nR_2}, n, P_e)$ code with $P_e \to 0$, the rate triple $(R_0, R_1, R_2)$ must satisfy (60)–(72) for some joint distribution $p(v_0)p(x_1|v_0)p(x_2|v_0)$.

Consider a $(2^{nR_0}, 2^{nR_1}, 2^{nR_2}, n, P_e)$ code with $P_e \to 0$. Note that $P_e \to 0$ implies $P_{e,1}^n \to 0$ and $P_{e,2}^n \to 0$. Applying Fano-inequality [22] on decoder 1, we obtain

$$H(W_0, W_1|Y_1^n) \leq n(R_0 + R_1)P_{e,1}^n + h(P_{e,1}^n) \triangleq n\epsilon_{1n}, \tag{74}$$

where $h(\cdot)$ is the binary entropy function, and $\epsilon_{1n} \to 0$ as $P_{e,1}^n \to 0$. It easily follows that

$$H(W_1|Y_1^n, W_0) \leq H(W_0, W_1|Y_1^n) \leq n\epsilon_{1n}. \tag{75}$$

By symmetry, we can also get

$$H(W_2|Y_2^n, W_0) \leq H(W_0, W_2|Y_2^n) \leq n\epsilon_{2n}. \tag{76}$$

We now expand the entropy term $H(Y_1^n, V_2^n|W_0, W_1)$ as

$$H(Y_1^n, V_2^n|W_0, W_1) \overset{(a)}{=} H(Y_1^n, V_2^n|X_1^n, W_0, W_1)$$

$$\overset{(b)}{=} H(V_2^n|X_1^n, W_0, W_1) + H(Y_1^n|V_2^n, X_1^n, W_0, W_1)$$

$$\overset{(c)}{=} H(Y_1^n|X_1^n, W_0, W_1) + H(V_2^n|Y_1^n, X_1^n, W_0, W_1),$$

where (a) follows from the fact that $X_1^n = f_1(W_0, W_1)$ is a deterministic function of $W_0$ and $W_1$ for a given $(2^{nR_0}, 2^{nR_1}, 2^{nR_2}, n, P_e)$ code; both (b) and (c) are based on the chain rule. Since $Y_1$ is a deterministic function of $X_1$ and $V_2$, $H(Y_1^n|V_2^n, X_1^n, W_0, W_1) = 0$. Similarly, due to $V_2 = h_1(Y_1, X_1)$, we have $H(V_2^n|Y_1^n, X_1^n, W_0, W_1) = 0$. Hence, we obtain the following equality

$$H(V_2^n|X_1^n, W_0, W_1) = H(Y_1^n|X_1^n, W_0, W_1),$$

which can be further simplified as follows

$$H(V_2^n|W_0, W_1) \overset{(a)}{=} H(Y_1^n|W_0, W_1),$$

$$H(V_2^n|W_0) \overset{(b)}{=} H(Y_1^n|W_0, W_1), \tag{77}$$

where (a) again follows from the deterministic relation between $X_1^n$ and $(W_0, W_1)$, and (b) follows from the conditional independence between $V_2^n$ and $W_1$ given $W_0$. Analogously, we can obtain

$$H(V_1^n|W_0) = H(Y_2^n|W_0 W_2). \tag{78}$$

One more pair of crucial inequalities are to be shown before we proceed to the main part of the converse. This pair are listed as follows

$$I(W_1; Y_1^n|W_0) \leq I(W_1; Y_1^n V_1^n|V_2^n W_0), \tag{79}$$

$$I(W_2; Y_2^n|W_0) \leq I(W_2; Y_2^n V_2^n|V_1^n W_0). \tag{80}$$

Inequality (79) can be derived as follows:

$$I(W_1; Y_1^n|W_0) = H(W_1|W_0) - H(W_1|Y_1^n W_0)$$

$$\overset{(a)}{\leq} H(W_1|V_2^n W_0) - H(W_1|Y_1^n V_2^n W_0)$$

$$\overset{(b)}{\leq} H(W_1|V_2^n W_0) - H(W_1|Y_1^n V_1^n V_2^n W_0)$$

$$= I(W_1; Y_1^n V_1^n|V_2^n W_0),$$

where (a) follows from the facts that $H(W_1|W_0) = H(W_1|V_2^n W_0)$ which is due to the conditional independence between $W_1$ and $V_2^n$ given $W_0$, and "conditioning reduces entropy", i.e., $H(W_1|Y_1^n V_2^n W_0) \leq H(W_1|Y_1^n W_0)$; and (b) follows from "conditioning reduces entropy" as well. Similarly, we can obtain (80).

Now we prove each of inequalities (60)–(72) with (75)–(80). Firstly, inequalities (60) and (61) are obvious.

For (62), we have

$$
\begin{aligned}
nR_1 = H(W_1) &= H(W_1|W_0) \\
&\stackrel{(a)}{=} H(W_1|W_0V_2^n) \\
&= I(W_1;Y_1^n|W_0V_2^n) + H(W_1|Y_1^nW_0V_2^n) \\
&\stackrel{(b)}{\leq} H(Y_1^n|W_0V_2^n) - H(Y_1^n|W_0W_1V_2^n) + n\epsilon_{1n} \\
&\stackrel{(c)}{=} H(Y_1^n|W_0V_2^n) + n\epsilon_{1n} \\
&\leq \sum_{i=1}^{n} H(Y_{1i}|V_{2i}W_0) + n\epsilon_{1n},
\end{aligned}
\tag{81}
$$

where (a) follows from the fact that $W_1$ and $V_2^n$ are conditionally independent given $W_0$; (b) follows from $H(W_1|Y_1^nW_0V_2^n) \leq H(W_1|Y_1^nW_0) \leq n\epsilon_{1n}$; (c) follows from $H(Y_1^n|W_0W_1V_2^n) = H(Y_1^n|X_1^nV_2^nW_0W_1) = 0$.

Analogously, for (63) we have

$$
nR_2 \leq \sum_{i=1}^{n} H(Y_{2i}|V_{1i}W_0) + n\epsilon_{2n}.
\tag{82}
$$

With respect to (64), we have

$$
\begin{aligned}
&n(R_1 + R_2) \\
&= H(W_1) + H(W_2) \\
&= H(W_1|W_0) + H(W_2|W_0) \\
&= I(W_1;Y_1^n|W_0) + H(W_1|Y_1^nW_0) + I(W_2;Y_2^n|W_0) + H(W_2|Y_2^nW_0) \\
&\stackrel{(a)}{\leq} H(Y_1^n|W_0) - H(Y_1^n|W_0W_1) + H(Y_2^n|W_0) - H(Y_2^n|W_0W_2) + n(\epsilon_{1n} + \epsilon_{2n}) \\
&\stackrel{(b)}{=} H(Y_1^n|W_0) - H(V_2^n|W_0) + H(Y_2^n|W_0) - H(V_1^n|W_0) + n(\epsilon_{1n} + \epsilon_{2n}) \\
&\leq H(Y_1^nV_1^n|W_0) - H(V_1^n|W_0) + H(Y_2^nV_2^n|W_0) - H(V_2^n|W_0) + n(\epsilon_{1n} + \epsilon_{2n}) \\
&= H(Y_1^n|V_1^nW_0) + H(Y_2^n|V_2^nW_0) + n(\epsilon_{1n} + \epsilon_{2n}) \\
&\leq \sum_{i=1}^{n} H(Y_{1i}|V_{1i}W_0) + \sum_{i=1}^{n} H(Y_{2i}|V_{2i}W_0) + n(\epsilon_{1n} + \epsilon_{2n}),
\end{aligned}
\tag{83}
$$

where (a) follows from inequalities (75) and (76); (b) follows from equalities (77) and (78).

Regarding to (65), we have

$$n(R_1 + R_2) = H(W_1|W_0) + H(W_2|W_0)$$

$$\overset{(a)}{\leq} I(W_1; Y_1^n|W_0) + I(W_2; Y_2^n|W_0) + n(\epsilon_{1n} + \epsilon_{2n})$$

$$\overset{(b)}{\leq} I(W_1; Y_1^n|W_0) + I(W_2; Y_2^n V_2^n|V_1^n W_0) + n(\epsilon_{1n} + \epsilon_{2n})$$

$$= I(W_1; Y_1^n|W_0) + I(W_2; V_2^n|V_1^n W_0) + I(W_2; Y_2^n|V_1^n V_2^n W_0) + n(\epsilon_{1n} + \epsilon_{2n})$$

$$\leq H(Y_1^n|W_0) - H(Y_1^n|W_0 W_1) + H(V_2^n|V_1^n W_0) - H(V_2^n|V_1^n W_2 W_0)$$
$$\quad + H(Y_2^n|V_1^n V_2^n W_0) - H(Y_2^n|V_1^n V_2^n W_2 W_0) + n(\epsilon_{1n} + \epsilon_{2n})$$

$$\overset{(c)}{=} H(Y_1^n|W_0) + H(Y_2^n|V_1^n V_2^n W_0) + n(\epsilon_{1n} + \epsilon_{2n})$$

$$\leq \sum_{i=1}^{n} H(Y_{1i}|W_0) + \sum_{i=1}^{n} H(Y_{2i}|V_{1i} V_{2i} W_0) + n(\epsilon_{1n} + \epsilon_{2n}), \tag{84}$$

where (a) follows from inequalities (75) and (76); (b) follows from inequality (79); (c) follows from the facts that 1) $H(Y_1^n|W_0 W_1) = H(V_2^n|V_1^n W_0)$, 2) $H(V_2^n|V_1^n W_2 W_0) = 0$ due to that $V_2^n$ is determined by $X_2^n$ which is again determined by $(W_0, W_2)$, and 3) $H(Y_2^n|V_1^n V_2^n W_2 W_0) = H(Y_2^n|X_2^n V_1^n V_2^n W_2 W_0) = 0$.

Similarly, we have

$$n(R_1 + R_2) \leq \sum_{i=1}^{n} H(Y_{2i}|W_0) + \sum_{i=1}^{n} H(Y_{1i}|V_{1i} V_{2i} W_0) + n(\epsilon_{1n} + \epsilon_{2n}), \tag{85}$$

which corresponds to (67).

For (66), we obtain

$$n(R_0 + R_1 + R_2)$$

$$= H(W_0 W_1) + H(W_2|W_0)$$

$$\overset{(a)}{\leq} I(W_0 W_1; Y_1^n) + I(W_2; Y_2^n|W_0) + n(\epsilon_{1n} + \epsilon_{2n})$$

$$\overset{(b)}{\leq} I(W_0 W_1; Y_1^n) + I(W_2; Y_2^n V_2^n|V_1^n W_0) + n(\epsilon_{1n} + \epsilon_{2n})$$

$$= I(W_0 W_1; Y_1^n) + I(W_2; V_2^n|V_1^n W_0) + I(W_2; Y_2^n|V_1^n V_2^n W_0) + n(\epsilon_{1n} + \epsilon_{2n})$$

$$\leq H(Y_1^n) - H(Y_1^n|W_0 W_1) + H(V_2^n|V_1^n W_0) - H(V_2^n|V_1^n W_2 W_0)$$
$$\quad + H(Y_2^n|V_1^n V_2^n W_0) - H(Y_2^n|V_1^n V_2^n W_2 W_0) + n(\epsilon_{1n} + \epsilon_{2n})$$

$$\overset{(c)}{=} H(Y_1^n) + H(Y_2^n|V_1^n V_2^n W_0) + n(\epsilon_{1n} + \epsilon_{2n})$$

$$\leq \sum_{i=1}^{n} H(Y_{1i}) + \sum_{i=1}^{n} H(Y_{2i}|V_{1i} V_{2i} W_0) + n(\epsilon_{1n} + \epsilon_{2n}), \tag{86}$$

where (a), (b) and (c) follow from the same arguments for (84). Note that the proof for (86) and the one for (84) only differ in the first few steps, and the rest follows from the same set of arguments and procedures.

Instead of expressing $n(R_0 + R_1 + R_2)$ as $H(W_0 W_1) + H(W_2| W_0)$, we set $n(R_0 + R_1 + R_2) = H(W_0|W_1) + H(W_0 W_2)$. Following the similar steps used in deriving (86), we can readily obtain

$$n(R_0 + R_1 + R_2) \leq \sum_{i=1}^{n} H(Y_{2i}) + \sum_{i=1}^{n} H(Y_{1i}|V_{1i}V_{2i}W_0) + n(\epsilon_{1n} + \epsilon_{2n}), \tag{87}$$

which corresponds to (68).

Now for (69), we can get

$$
\begin{aligned}
&n(2R_1 + R_2) \\
&= H(W_1|W_0) + H(W_1|W_0) + H(W_2|W_0) \\
&\overset{(a)}{\leq} I(W_1; Y_1^n|W_0) + I(W_1; Y_1^n|W_0) + I(W_2; Y_2^n|W_0) + n(2\epsilon_{1n} + \epsilon_{2n}) \\
&\overset{(b)}{\leq} I(W_1; Y_1^n|W_0) + I(W_1; Y_1^n V_1^n|V_2^n W_0) + I(W_2; Y_2^n|W_0) + n(2\epsilon_{1n} + \epsilon_{2n}) \\
&= I(W_1; Y_1^n|W_0) + I(W_1; V_1^n|V_2^n W_0) + I(W_1; Y_1^n|V_1^n V_2^n W_0) \\
&\quad + I(W_2; Y_2^n|W_0) + n(2\epsilon_{1n} + \epsilon_{2n}) \\
&= H(Y_1^n|W_0) - H(Y_1^n|W_0 W_1) + H(V_1^n|V_2^n W_0) - H(V_1^n|V_2^n W_0 W_1) \\
&\quad + H(Y_1^n|V_1^n V_2^n W_0) - H(Y_1^n|V_1^n V_2^n W_0 W_1) + H(Y_2^n|W_0) \\
&\quad - H(Y_2^n|W_0 W_2) + n(2\epsilon_{1n} + \epsilon_{2n}) \\
&\overset{(c)}{=} H(Y_1^n|W_0) - H(Y_1^n|W_0 W_1) + H(Y_1^n|V_1^n V_2^n W_0) + H(Y_2^n|W_0) + n(2\epsilon_{1n} + \epsilon_{2n}) \\
&\overset{(d)}{=} H(Y_1^n|W_0) - H(V_2^n|W_0) + H(Y_1^n|V_1^n V_2^n W_0) + H(Y_2^n|W_0) + n(2\epsilon_{1n} + \epsilon_{2n}) \\
&\leq H(Y_1^n|W_0) - H(V_2^n|W_0) + H(Y_1^n|V_1^n V_2^n W_0) + H(Y_2^n V_2^n|W_0) + n(2\epsilon_{1n} + \epsilon_{2n}) \\
&= H(Y_1^n|W_0) + H(Y_1^n|V_1^n V_2^n W_0) + H(Y_2^n|V_2^n W_0) + n(2\epsilon_{1n} + \epsilon_{2n}) \\
&\leq \sum_{i=1}^{n} H(Y_{1i}|W_0) + \sum_{i=1}^{n} H(Y_{1i}|V_{1i}V_{2i}W_0) + \sum_{i=1}^{n} H(Y_{2i}|V_{2i}W_0) + n(2\epsilon_{1n} + \epsilon_{2n}), \tag{88}
\end{aligned}
$$

where (a) follows from inequalities (75) and (76); (b) follows from inequality (79); (c) follows from the facts that $H(V_1^n|V_2^n W_0) = H(V_1^n|W_0) = H(Y_2^n|W_0 W_2)$, $H(V_1^n|V_2^n W_0 W_1) = H(V_1^n|X_1^n V_2^n W_0 W_1) = 0$ and $H(Y_1^n|V_1^n V_2^n W_0 W_1) = H(Y_1^n|V_1^n X_1^n V_2^n W_0 W_1) = 0$; (d) follows from $H(V_2^n|W_0) = H(Y_1^n|W_0 W_1)$. Following similar procedures, we can easily obtain

$$
\begin{aligned}
n(R_1 + 2R_2) &\leq \sum_{i=1}^{n} H(Y_{2i}|W_0) + \sum_{i=1}^{n} H(Y_{2i}|V_{1i}V_{2i}W_0) + \sum_{i=1}^{n} H(Y_{1i}|V_{1i}W_0) \\
&\quad + n(\epsilon_{1n} + 2\epsilon_{2n}), \tag{89}
\end{aligned}
$$

$$
\begin{aligned}
n(R_0 + 2R_1 + R_2) &\leq \sum_{i=1}^{n} H(Y_{1i}) + \sum_{i=1}^{n} H(Y_{1i}|V_{1i}V_{2i}W_0) + \sum_{i=1}^{n} H(Y_{2i}|V_{2i}W_0) \\
&\quad + n(2\epsilon_{1n} + \epsilon_{2n}), \text{ and} \tag{90}
\end{aligned}
$$

$$n(R_0 + R_1 + 2R_2) \leq \sum_{i=1}^{n} H(Y_{2i}) + \sum_{i=1}^{n} H(Y_{2i}|V_{1i}V_{2i}W_0) + \sum_{i=1}^{n} H(Y_{1i}|V_{1i}W_0)$$
$$+ n(\epsilon_{1n} + 2\epsilon_{2n}), \tag{91}$$

which correspond to (71), (70) and (72) respectively.

Note that we have derived a number of inequalities (81)–(91) which, together with (60) and (61), upper bound the rate triple $(R_0, R_1, R_2)$ of the given code for the DICC channel. We now adopt the technique which was used to prove the converse of the capacity region of the MACC in [19] and [20]. Define $V_0 = W_0$, or equivalently $V_{0i} = W_0$, i.e., $V_0$ or $V_{0i}$ is an auxiliary random variable uniformly distributed over the common message set $\mathcal{W}_0 = \{1, ..., M_0\}$. Since $X_1$ and $X_2$ are conditionally independent given $W_0$, i.e., $p(x_{1i}, x_{2i}|w_0) = p(x_{1i}|w_0)p(x_{2i}|w_0)$, we can write

$$p(x_{1i}, x_{2i}|v_{0i}) = p(x_{1i}|v_{0i})p(x_{2i}|v_{0i}).$$

Note that due to the introduction of $V_0$, the region inherits the convexity from the achievable rate region for the general ICC. We can now conclude that as $n \to \infty$ and $P_e \to 0$, the rate of the given code $(R_0, R_1, R_2)$ is bounded by (60)–(72) for some choice of joint distribution $p(v_0)p(x_1|v_0)p(x_2|v_0)$. This completes the proof of the converse and the theorem. ∎

*Remark 15:* 1) As mentioned earlier, our assumption of this class of deterministic channel is slightly different from the one given in [11]. We directly require the existence of functions $V_2 = h_1(Y_1, X_1)$ and $V_1 = h_2(Y_2, X_2)$ such that we have the two equalities $H(V_2^n|W_0) = H(Y_1^n|W_0W_1)$ and $H(V_1^n|W_0) = H(Y_2^n|W_0W_2)$. As demonstrated in the above proof, the two inequalities are crucial, without which we are not able to establish the converse. Moreover, the two equalities in fact reduce to the assumptions made in [11] in the absence of common information. Therefore, we can claim that the existence of $V_2 = h_1(Y_1, X_1)$ and $V_1 = h_2(Y_2, X_2)$ is the more general condition for this class of deterministic interference channels. 2) The capacity region of the class of DICCs derived above generalizes the one given in [11].

## VI. Conclusions

In this paper, we have investigated the general discrete memoryless interference channel with common information, and obtained an achievable rate region for the channel by applying a random coding scheme consisting of the generalized successive superposition encoding and simultaneous decoding. The achievable rate region is found to be potentially tight, as it not only generalizes some important existing results for the interference channel with or without common information, i.e., the capacity region of the strong interference channel with common information and the Chong-Motani-Garg region (one of the two best achievable rate regions for the interference channel without common information) are shown as special cases of our achievable rate region; but also is shown to be the exact capacity region for a class of deterministic interference channels with common information. Nevertheless, it remains a challenge to establish a converse to our achievable rate region for the general discrete memoryless interference channel with common information.

## Appendix
### Proof of Convexity $\mathcal{R}_m$

Let $(R_0^1, R_{12}^1, R_{11}^1, R_{21}^1, R_{22}^1)$ and $(R_0^2, R_{12}^2, R_{11}^2, R_{21}^2, R_{22}^2)$ be two arbitrary rate quintuples belonging to $\mathcal{R}_m$. It suffices to show that for given any $\alpha \in [0, 1]$, we have $(\alpha R_0^1 + (1 -$ $\alpha)R_0^2, \alpha R_{12}^1 + (1-\alpha)R_{12}^2, \alpha R_{11}^1 + (1-\alpha)R_{11}^2, \alpha R_{21}^1 + (1-\alpha)R_{21}^2, \alpha R_{22}^1 + (1-\alpha)R_{22}^2) \in \mathcal{R}_m$. Note that the rate region $\mathcal{R}_m$ is the union of regions $\mathcal{R}_m(p)$ over all $p(\cdot) \in \mathcal{P}^*$. Thus, there must exist two sets of auxiliary random variables $(U_0^1, U_1^1, U_2^1)$ and $(U_0^2, U_1^2, U_2^2)$ such that their joint distributions $p_1(\cdot)$ and $p_2(\cdot)$ factor as

$$p_1(u_0^1, u_1^1, u_2^1, x_1, x_2, y_1, y_2) = p(u_0^1)p(u_1^1|u_0^1)p(u_2^1|u_0^1)p(x_1|u_1^1, u_0^1)p(x_2|u_2^1, u_0^1)p(y_1, y_2|x_1, x_2),$$

$$p_2(u_0^2, u_1^2, u_2^2, x_1, x_2, y_1, y_2) = p(u_0^2)p(u_1^2|u_0^2)p(u_2^2|u_0^2)p(x_1|u_1^2, u_0^2)p(x_2|u_2^2, u_0^2)p(y_1, y_2|x_1, x_2).$$

Let $T$ be the independent random variable, taking the value 1 with probability $\alpha$ and 2 with probability $1-\alpha$. We define a new set of auxiliary random variables $(U_0, U_1, U_2)$ such that $U_0 = (U_0^T, T)$, $U_1 = U_1^T$, and $U_2 = U_2^T$, and then their joint distribution $p_3(\cdot)$ can factor

$$p_3(u_0, u_1, u_2, x_1, x_2, y_1, y_2) = p(u_0)p(u_1|u_0)p(u_2|u_0)p(x_1|u_1, u_0)p(x_2|u_2, u_0)p(y_1, y_2|x_1, x_2).$$

Since $p_3(\cdot) \in \mathcal{P}^*$, we have $\mathcal{R}_m(p_3) \subseteq \mathcal{R}_m$. It is easy to show that $(\alpha R_0^1 + (1-\alpha)R_0^2, \alpha R_{12}^1 + (1-\alpha)R_{12}^2, \alpha R_{11}^1 + (1-\alpha)R_{11}^2, \alpha R_{21}^1 + (1-\alpha)R_{21}^2, \alpha R_{22}^1 + (1-\alpha)R_{22}^2) \in \mathcal{R}_m(p_3)$ by following the steps used to prove the convexity of the capacity region for the MACC in the Appendix A of [20]. Therefore, we conclude that $(\alpha R_0^1 + (1-\alpha)R_0^2, \alpha R_{12}^1 + (1-\alpha)R_{12}^2, \alpha R_{11}^1 + (1-\alpha)R_{11}^2, \alpha R_{21}^1 + (1-\alpha)R_{21}^2, \alpha R_{22}^1 + (1-\alpha)R_{22}^2) \in \mathcal{R}_m(p_3) \subseteq \mathcal{R}_m$, which proves the convexity of $\mathcal{R}_m$.

## REFERENCES

[1] C. E. Shannon, "Two-way communication channels," in *Proc. 4th Berkeley Symp. on Mathematical Statistics and Probability*, vol. 1, Berkeley, CA, 1961, pp. 611–644.

[2] R. Ahlswede, "The capacity region of a channel with two senders and two receivers," *Annals of Probability*, vol. 2, no. 5, pp. 805–814, 1974.

[3] A. B. Carleial, "A case where interference does not reduce capacity," *IEEE Trans. Inform. Theory*, vol. IT-21, no. 5, pp. 569–570, Sept. 1975.

[4] H. Sato, "Two-user communication channels," *IEEE Trans. Inform. Theory*, vol. IT-23, no. 3, pp. 295–304, May 1977.

[5] ——, "On degraded Gaussian two-user channels," *IEEE Trans. Inform. Theory*, vol. IT-24, no. 5, pp. 637–640, May 1978.

[6] ——, "On the capacity region of a discrete two-user channel for strong interference," *IEEE Trans. Inform. Theory*, vol. IT-24, no. 3, pp. 377–379, May 1978.

[7] A. B. Carleial, "Interference channels," *IEEE Trans. Inform. Theory*, vol. IT-24, no. 1, pp. 60–70, Jan. 1978.

[8] R. Benzel, "The capacity region of a class of discrete additive degraded interference channels," *IEEE Trans. Inform. Theory*, vol. IT-25, no. 2, pp. 228–231, Mar. 1979.

[9] T. S. Han and K. Kobayashi, "A new achievable rate region for the interference channel," *IEEE Trans. Inform. Theory*, vol. IT-27, no. 1, pp. 49–60, Jan. 1981.

[10] H. Sato, "The capacity of the Gaussian interference channel under strong interference," *IEEE Trans. Inform. Theory*, vol. IT-27, no. 6, pp. 786–788, Nov. 1981.

[11] A. A. E. Gamal and M. H. M. Costa, "The capacity region of a class of deterministic interference channels," *IEEE Trans. Inform. Theory*, vol. IT-28, no. 2, pp. 343–346, Mar. 1982.

[12] M. H. M. Costa and A. A. E. Gamal, "The capacity region of the discrete memoryless interference channel with strong interference," *IEEE Trans. Inform. Theory*, vol. 35, no. 5, pp. 710–711, Sept. 1987.

[13] I. Sason, "On achievable rate regions for the Gaussian interference channel," *IEEE Trans. Inform. Theory*, vol. 50, no. 6, pp. 1345–1356, June 2004.

[14] G. Kramer, "Genie-aided outer bounds on the capacity of interference channels," *IEEE Trans. Inform. Theory*, vol. 50, no. 3, pp. 581–586, Mar. 2004.

[15] H. F. Chong, M. Motani, and H. K. Garg, "A comparison of two achievable rate regions for the interference channel," in *ITA Workshop*, San Diego, Feb. 2006.

[16] H. F. Chong, M. Motani, H. K. Garg, and H. E. Gamal, "On a simplification of the Han-Kobayashi rate region for the interference channel," *IEEE Trans. Inform. Theory*, submitted for publication.

[17] T. M. Cover, "An achievable rate region for the broadcast channel," *IEEE Trans. Inform. Theory*, vol. IT-21, pp. 399–404, July 1975.

[18] I. Maric, R. Yates, and G. Kramer, "The capacity region of the strong interference channel with common information," in *Proc. Asilomar Conf. on Signals, Systems, and Computers*, Pacific Grove, CA, USA, Oct. 30–Nov. 2, 2005, pp. 1737–1741.

[19] D. Slepian and J. K. Wolf, "A coding theorem for multiple access channels with correlated sources," *Bell System Tech. J.*, vol. 52, pp. 1037–1076, Sept. 1973.

[20] F. M. J. Willems, "Information theoretical results for the discrete memoryless multiple access channel," in *Ph.D. dissertation*, Katholieke Universiteit Leuven, Belgium, Oct. 1982.

[21] G. Kramer, "Review of rate regions for interference channels," in *IZS Workshop*, Zurich, Feb. 2006.

[22] T. M. Cover and J. A. Thomas, *Elements of Information Theory*. New York: John Wiley & Sons, 1994.

[23] F. M. Willems, "The multiple-access channel with cribbing encoders revisited," in *MSRI Workshop Mathematics of Relaying and Cooperation in Communication Networks*, Berkeley, April 10-12, 2006.